\documentclass[11pt]{article}

		\usepackage{amsmath}
		\usepackage{amssymb}
		\usepackage{amsthm}
		\usepackage{mathrsfs}
		\usepackage{esvect}
		\usepackage{latexsym}
		\usepackage{mathtools}
		\usepackage{graphicx}
		\usepackage{epstopdf}
		\usepackage{float}
		\usepackage{subfig}

		\usepackage[utf8]{inputenc}
		\usepackage{wasysym}
		\usepackage[cam,letter,center]{crop}
		\usepackage[margin=1.5in,inner=1.6in,outer=1.4in]{geometry}
		\usepackage{fancyhdr}
		\usepackage{textgreek}
		\DeclareMathSymbol{\Ohm}{\mathalpha}{operators}{10}

		\usepackage{sectsty}
		\allsectionsfont{\large}
		\sectionfont{}
		\subsectionfont{}
		\subsubsectionfont{\vspace{-0.5\baselineskip}\bfseries}
		\makeatletter
		\def\@seccntformat#1{\@ifundefined{#1@cntformat}{\csname the#1\endcsname\quad}{\csname #1@cntformat\endcsname}}
		\def\section@cntformat{\thesection.\quad}
		\def\subsection@cntformat{\thesubsection.\quad}
		\makeatother
		\usepackage{titlesec}
		\titlespacing*{\subsection}{0pt}{0.7\baselineskip}{\baselineskip}

		\usepackage{hyphenat}
		\catcode`@=11
		\MakeRobust{\vphantom}
		\usepackage{stackengine}[2013-10-15]
		\usepackage{scalerel}
		
		\renewcommand{\sin}[1]{\text{sin}\!\left({#1}\right)}
		\newcommand{\sinn}[2]{\text{sin}^{#1}\!\left({#2}\right)}
		\renewcommand{\cos}[1]{\text{cos}\!\left({#1}\right)}

		\renewcommand{\arcsin}[1]{\text{arcsin}\!\left({#1}\right)}

		\renewcommand{\log}[1]{\text{ln}\!\left({#1}\right)}
		\newcommand{\logg}[2]{\text{log}_{#1}\!\left({#2}\right)}
		
		\newcommand{\Real}[1]{\text{Re}\!\left({#1}\right)}

		\newcommand{\Imag}[1]{\text{Im}\!\left({#1}\right)}

		\renewcommand{\arg}[1]{\text{arg}\!\left({#1}\right)}
		
		\newcommand{\pder}[2]{\frac{\partial{#1}}{\partial{#2}}}
		
		\newcommand{\integral}[4]{\int_{#1}^{#2}{#3}\,\text{d}{#4}}
		
		\DeclareMathOperator*{\stksum}{\scalerel*{\Xi}{\sum}}
		
		\usepackage[makeroom]{cancel}
		\renewcommand{\parallel}{{\,\scalebox{0.6}{\raisebox{0.8ex}{\cancel{\vphantom{.}\phantom{\,}}\cancel{\vphantom{.}\phantom{\,}}}}}}
		\renewcommand{\leq}{\leqslant}
		
		\renewcommand{\.}{\,\!}
		\newcommand*{\eq}[1]{\begin{eqnarray}#1\end{eqnarray}}
		\newcommand{\tab}{\hskip\arraycolsep\,}

		\newcommand{\figwidth}{0.85\linewidth}
		
		\newcommand{\graphwidth}{1.2*\figwidth/sqrt(2)}
		\newcommand{\graphheight}{2*\graphwidth/(1+sqrt(5))}
		\usepackage[font=small]{caption}
		\usepackage{boxhandler}
		\captionStyle{n}{l}
		\usepackage{pgfplots}
		\pgfplotsset{compat=1.9}
		\usepgfplotslibrary{fillbetween}
		\definecolor{darkgreen}{rgb}{0,0.8,0}
		\definecolor{darkred}{rgb}{0.6,0,0}
		\definecolor{grey50}{rgb}{0.5,0.5,0.5}
		\definecolor{grey80}{rgb}{0.8,0.8,0.8}
		\definecolor{lightblue}{rgb}{0.3,0.3,1}
		\definecolor{orangee}{rgb}{1,0.45,0.19}
		\definecolor{yelloww}{rgb}{0.97,0.86,0.15}
		\newcommand*{\fig}[4]{\begin{figure}[b!]\begin{center}{\hrule\ \\\ \\\includegraphics[width=#2]{#1}}\captionsetup{singlelinecheck=off}\caption[.]{#3}\label{#4}\end{center}\end{figure}}

		\usepackage{array}
		\newcolumntype{L}[1]{>{\raggedright\let\newline\\\arraybackslash\hspace{0pt}}m{#1}}
		\newcolumntype{C}[1]{>{\centering\let\newline\\\arraybackslash\hspace{0pt}}m{#1}}
		\newcolumntype{R}[1]{>{\raggedleft\let\newline\\\arraybackslash\hspace{0pt}}m{#1}}
		\usepackage{multirow}
		\usepackage{multicol}
		\usepackage{tabularx,colortbl}
		\numberwithin{table}{section}

		\usepackage{verbatim}
		\usepackage{fancyvrb}

		\usepackage[super,comma,sort&compress]{natbib}
		
		\makeatletter
		\renewcommand\@biblabel[1]{#1.}
		\makeatother
		\usepackage{xifthen}
		\newcommand{\article}[6]{#1 (#2): \textit{#3}, #4 #5, \mbox{#6}}

		\newcommand{\book}[6]{#1 (#2): \textit{#3}, \ifthenelse{\isempty{#4}}{}{\ifthenelse{\equal{\detokenize{#4}}{\detokenize{2}}}{second}{\ifthenelse{\equal{\detokenize{#4}}{\detokenize{3}}}{third}{\ifthenelse{\equal{\detokenize{#4}}{\detokenize{4}}}{fourth}{\ifthenelse{\equal{\detokenize{#4}}{\detokenize{5}}}{fifth}{\ifthenelse{\equal{\detokenize{#4}}{\detokenize{6}}}{sixth}{\ifthenelse{\equal{\detokenize{#4}}{\detokenize{7}}}{seventh}{\ifthenelse{\equal{\detokenize{#4}}{\detokenize{8}}}{eighth}{\ifthenelse{\equal{\detokenize{#4}}{\detokenize{9}}}{ninth}{#4}}}}}}}} edition, }#5, \mbox{#6}}
		\newcommand{\bookarticle}[7]{#1 (#2): \textit{#3}, in \textit{#4}\ifthenelse{\isempty{#5}}{}{ #5}, #6, \mbox{#7}}

		\newcommand{\'}{\'\i}

		\newcommand{\Es}{E_\text{s}}

		\renewcommand{\ng}{n_\text{g}}
		
		\newcommand{\ns}{n_\text{s}}
		\newcommand{\nw}{n_\text{w}}

		\newcommand{\rgs}{r_\text{gs}}
		\newcommand{\rgw}{r_\text{gw}}

		\newcommand{\rsg}{r_\text{sg}}
		\newcommand{\rsw}{r_\text{sw}}

		\newcommand{\tgs}{t_\text{gs}}

		\newcommand{\tsg}{t_\text{sg}}

		\newcommand{\thetac}{\theta_\text{c}}
		\newcommand{\thetamax}{\theta_\text{max}}
		\newcommand{\thetas}{\theta_\text{s}}

\begin{document}\sloppy

\title{A model for the complex reflection coefficient of a collection of parallel layers}
\author{\underline{Alexander Nahmad-Rohen} (Nahmad-RohenA@cardiff.ac.uk),\\Wolfgang Langbein}
\date{}
\maketitle

Reflectometry is a technique that uses the light reflected by a sample to determine properties of the sample. Interferometric reflectometry uses interference between two beams, one of which is incident on ---and reflected back by--- a sample and one of which is not, to obtain the complex electric field rather than merely its intensity. Since this interference allows one to retrieve an increased amount of information about the light, it also allows one to obtain more information about the sample, such as a thin layer. We will apply the methods derived here to the case of a planar lipid bilayer.

\section{Reflection by a thin layer}\label{sec-s}

Suppose a sample consists of a thin layer of a homogeneous material of thickness $d$ and refractive index $\ns$ which does not cover the entirety of a flat glass surface of refractive index $\ng$ and is submerged in water, which has a refractive index $\nw$ (figure~\ref{fig-thinsample1}).

For normal incidence, the Fresnel reflection coefficient at the interface between a material with refractive index $n_j$ (through which a light beam travels) and a material with refractive index $n_k$ (which reflects the beam) is
\eq{r_{jk} & = & \frac{n_j-n_k}{n_j+n_k}\nonumber}
and the Fresnel transmission coefficient at that interface is
\eq{t_{jk} & = & \frac{2n_j}{n_j+n_k}.\nonumber}

The reflection coefficient of a region of the sample where there is no material between the glass and the water (figure~\ref{fig-thinsample1}, left side), then, is simply
\eq{\rgw & = & \frac{\ng-\nw}{\ng+\nw}.\nonumber}
For a region where there is a layer of material (figure~\ref{fig-thinsample1}, right side), some of the light will be reflected at the glass-material interface and some of it will be transmitted. The transmitted light might then be transmitted at the material-water interface, or it may be reflected any number $\ell$ of times at said interface and either be reflected $\ell-1$ times at the material-glass interface and eventually transmitted back through this interface or be reflected $\ell$ times at the material-glass interface and eventually transmitted through the material-water interface. For reflection, we are interested in the first case only. Therefore, the reflection coefficient of such a region is
\eq{s & = & \rgs+\tgs\tsg\sum_{\ell=1}^\infty\rsg\.^{\ell-1}\rsw\.^\ell e^{2\ell ikd\ns} \tab = \tab \rgs+\frac{\tgs\tsg}{\rsg}\,\sum_{\ell=1}^\infty\left(\rsg\rsw e^{2ikd\ns}\right)^\ell,\nonumber}
where the first term corresponds to reflection at the glass-material interface and the exponential in the sum is due to the fact that light reflected $\ell$ times at the material-water interface and $\ell-1$ times at the material-glass interface travels $2\ell$ times through the material. Here, $k$ is the wave vector of the light in vacuum. Because $|\rsg\rsw e^{2ikd\ns}|<1$, this is equal to
\eq{s & = & \rgs+\frac{\tgs\rsw\tsg e^{2ikd\ns}}{1-\rsg\rsw e^{2ikd\ns}}\nonumber\\
& = & \frac{(\ng+\ns)(\ns-\nw)e^{2ikd\ns}+(\ng-\ns)(\ns+\nw)}{(\ng-\ns)(\ns-\nw)e^{2ikd\ns}+(\ng+\ns)(\ns+\nw)}.\label{eq-s1}}
If $d=0$ or $\ns=\nw$, this reduces to $\rgw$. If $\ns=\ng$, it instead reduces to $e^{2ikd\ng}\rgw$ due to the fact that the light must still travel an additional distance $2d$ through material with refractive index $\ng$.

\fig{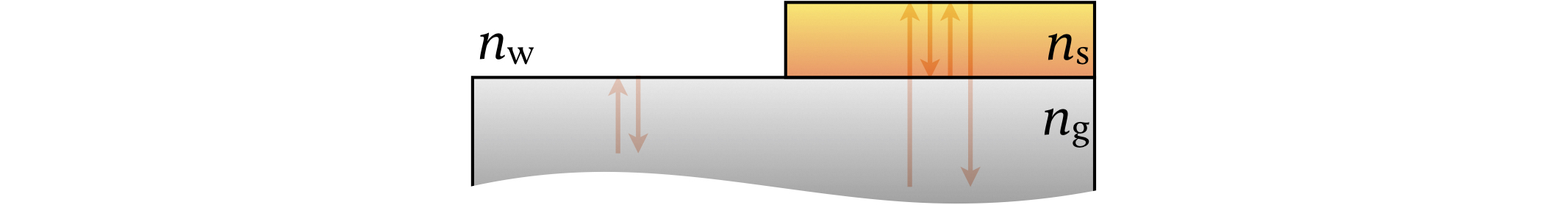}{\figwidth}{Reflection from a thin sample.}{fig-thinsample1}

\begin{figure}[b!]
\begin{center}
\hrule\ \\\ \\
\begin{minipage}[m]{\figwidth}
\qquad\includegraphics[width=\figwidth]{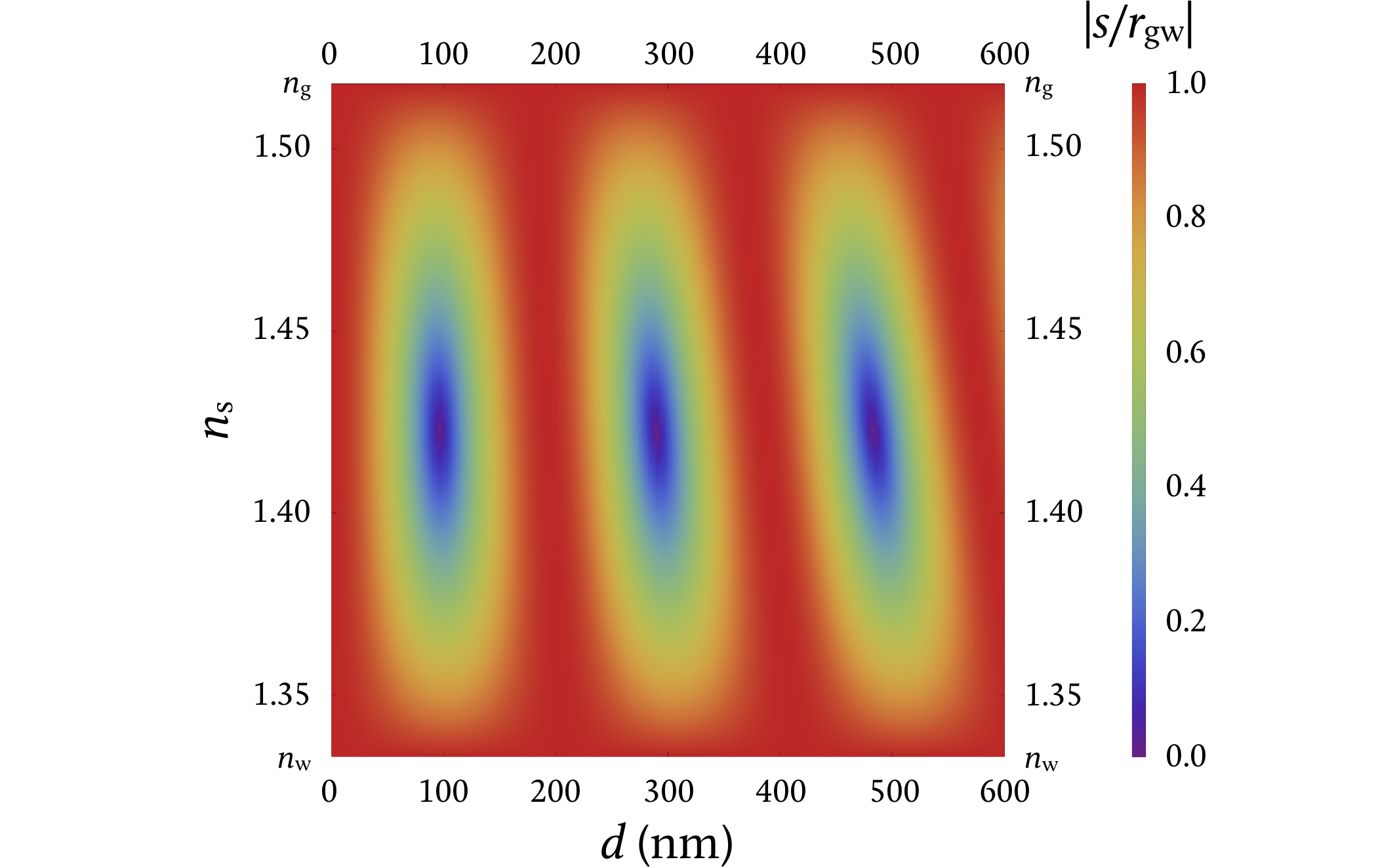}
\end{minipage}\\[2.5ex]
\begin{minipage}[m]{\figwidth}
\qquad\includegraphics[width=\figwidth]{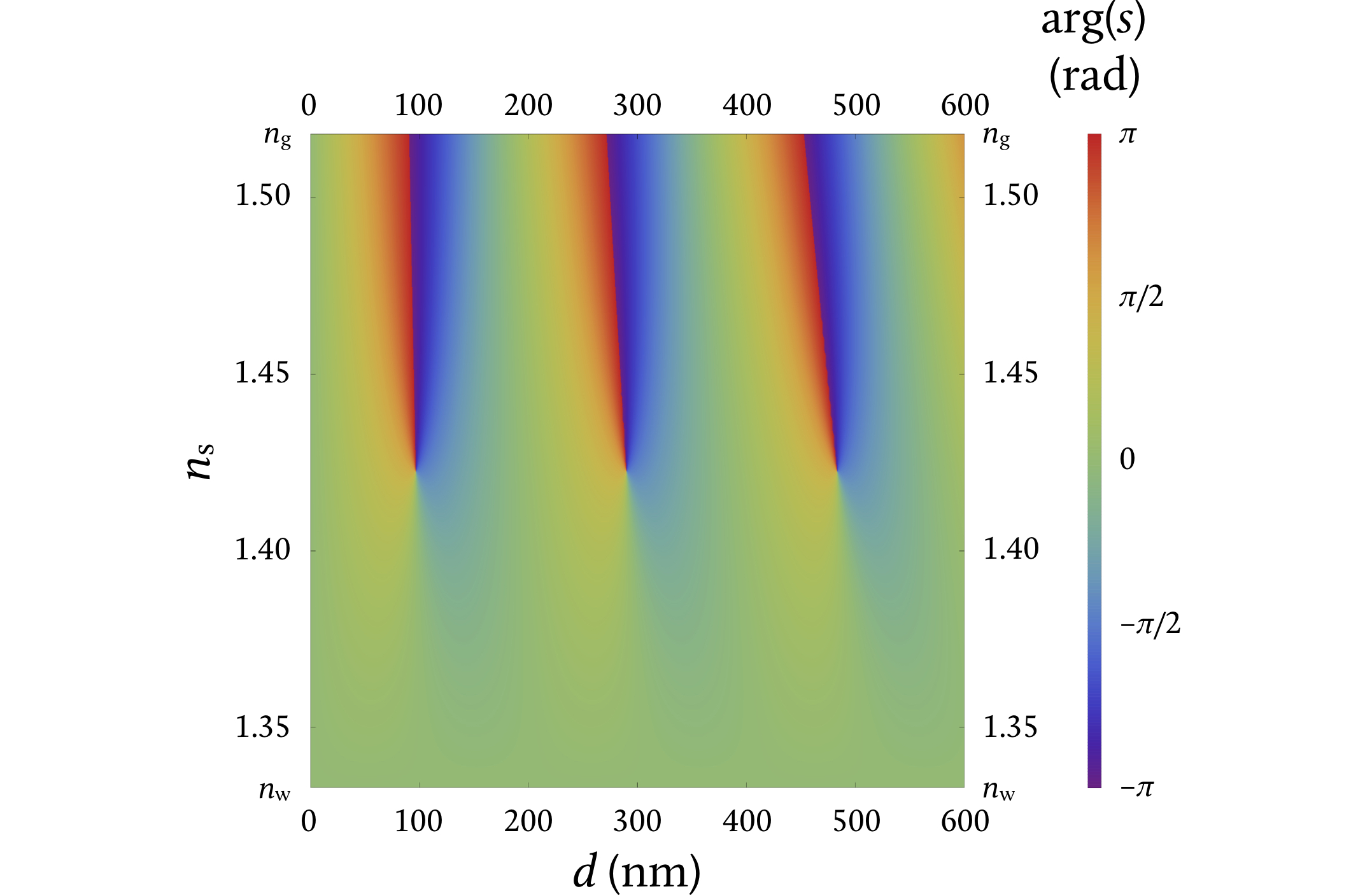}
\end{minipage}
\caption{Density graphs of $|s/\rgw|$ (top) and $\arg{s}$ (bottom) as functions of $d$ and $\ns$ for normal incidence with $\nw=1.333$, $\ng=1.518$ and $\lambda=550$~nm.}\label{fig-1-000}
\end{center}
\end{figure}

Here, we have assumed that the sample does not change the polarisation of the beam (i.e.~it is not birefringent) and thus $s$ is a scalar. If it has in-plane birefringence, $s$ is instead a $2\times2$ matrix and equation~\ref{eq-s1} is no longer appropriate; if it presents out-of-plane birefringence, the calculation of $s$ can become truly complicated. It should be noted that the type of sample we are interested in here, a lipid bilayer, consists of a $2$-dimensional array of lipid molecules oriented approximately perpendicular to said array and can thus be considered approximately isotropic for light travelling parallel (or approximately parallel) to the molecules; indeed, the birefringence of a lipid bilayer is negligible for our purposes.\cite{ref-ShawBB101} We will therefore ignore the effects of birefringence in what follows.

\begin{figure}[b!]
\begin{center}
\hrule\ \\\ \\
\begin{minipage}[m]{\figwidth}
\qquad\includegraphics[width=\figwidth]{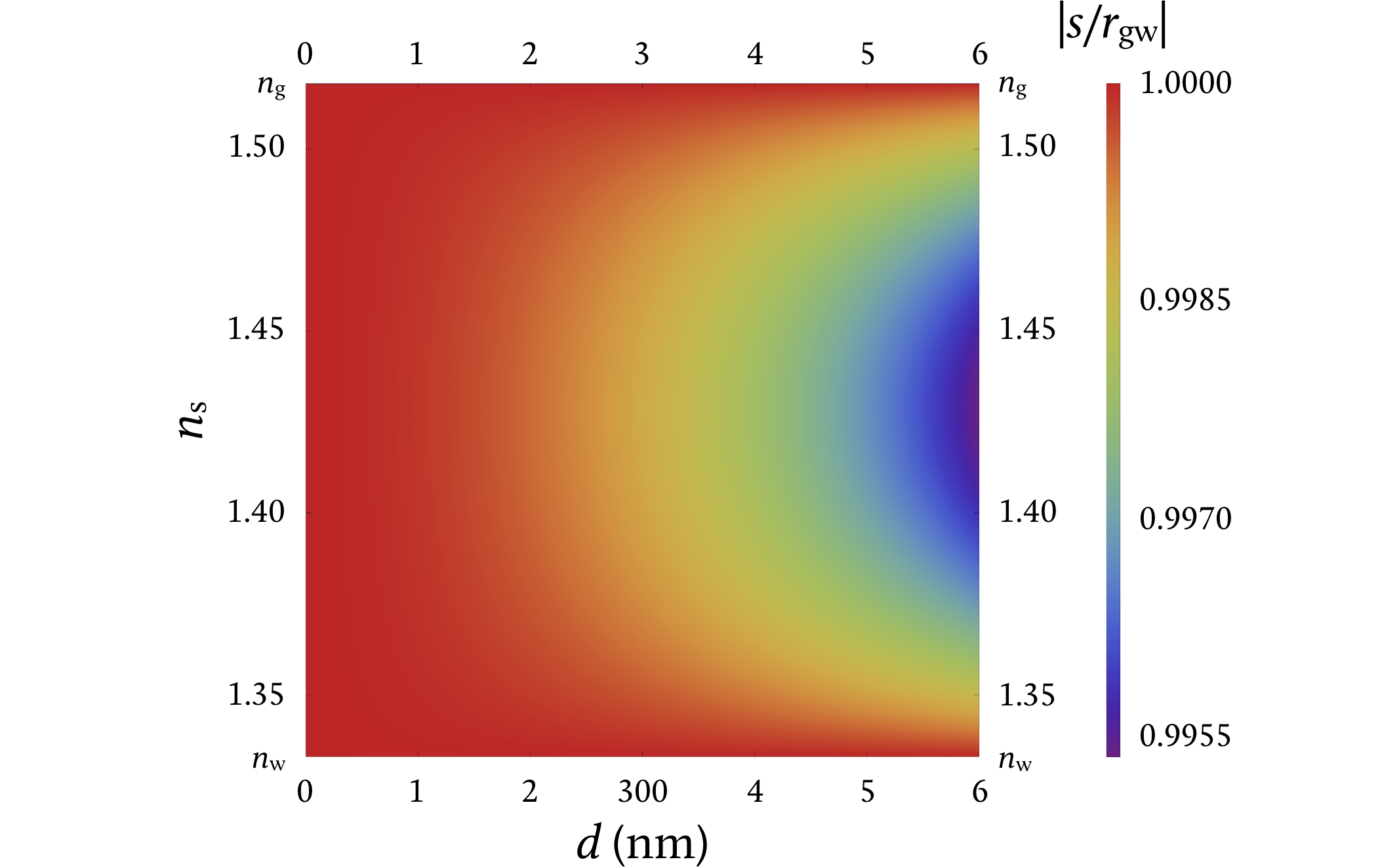}
\end{minipage}\\[2.5ex]
\begin{minipage}[m]{\figwidth}
\qquad\includegraphics[width=\figwidth]{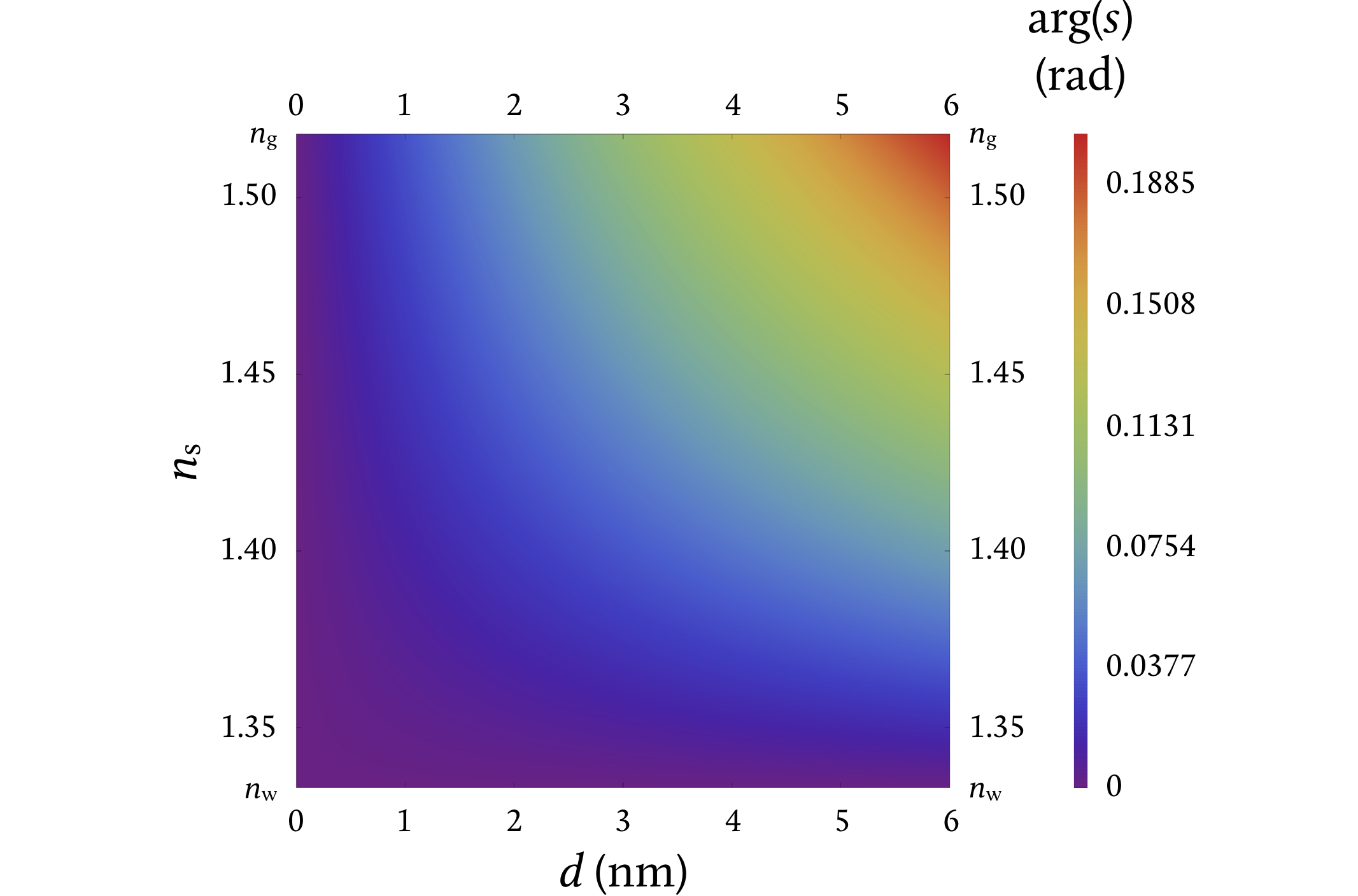}
\end{minipage}
\caption{Density graphs of $|s/\rgw|$ (top) and $\arg{s}$ (bottom) as functions of $d$ and $\ns$ for normal incidence with a reduced $d$ range and with $\nw=1.333$, $\ng=1.518$ and $\lambda=550$~nm.}\label{fig-1-000-red}
\end{center}
\end{figure}

Figure~\ref{fig-1-000} shows graphs of $|s|$ and $\arg{s}$ for $d$ between $0$~nm and $600$~nm, $\nw=1.333\leq\ns\leq1.518=\ng$, and $\lambda=2\pi/k=550$~nm. These values for the refractive indices were chosen because they correspond to water\cite{ref-HechtO} and the glass which microscope slides and coverslips are typically made of;\cite{ref-OcheiMLS} the refractive index of a lipid bilayer typically falls between these values.\cite{ref-HowlandBJ92,ref-DevanathanFEBSJ273} The thickness range was chosen to show three periods of $s$. It is interesting to note the existence of a value of $\ns$ below which $\arg{s}$ may only take values between $-\pi/2$ and $\pi/2$, meaning the reflection coefficient has a positive real part; this value will be calculated in section~\ref{sec-rphi}. Note also that, for $\nw\leq\ns\leq\ng$, we have $|s|\leq\rgw$ regardless of the value of $d$, meaning that the presence of the layer either reduces the amount of reflected light (by spatially distributing the refractive index step from $\ng$ to $\nw$ and giving rise to an interference which is not fully constructive) or does not affect it.

Figure~\ref{fig-1-000-red} shows the same graphs as figure~\ref{fig-1-000}, but for $d$ between $0$~nm and $6$~nm only. This range of $d$ corresponds to the thickness one would expect from a lipid bilayer, which is about $4$~nm thick. Note that $|s|$ changes very little in this region --- the difference between no layer and a $6$-nm layer is only about $0.5$\% for $\ns\approx1.425$ and even less for other values of $\ns$. This is due to the fact that the sample is very thin; a small value of $d$ will result in a small value of $2kd\ns$, which in turn means that the first few reflections of the beam within the sample interfere mostly constructively; by the time the number of reflections is large enough for the interference to be destructive, the amplitude of the beam is so small (due to the fact that $\rsw,\rsg<1$) that it contributes very little to the reflected field $\Es$. The variation in $\arg{s}$ is also reduced, but much less so --- it is a few percent even for intermediate values of $\ns$. In fact, if we only take the first reflection into account, the reflection coefficient becomes
\eq{s & \approx & \rgs+\tgs\rsw\tsg e^{2ikd\ns};\nonumber}
the difference between this and the complete reflection coefficient given by equation~\ref{eq-s1} is less than $0.06$\% throughout the range considered, as might be expected by noting that the denominator of the second term in the first line of equation~\ref{eq-s1} is approximately equal to $1$ because $\rsg\rsw<\rgw^2\approx0.0042$. Thus, taking only one reflection into account is an acceptable approximation.

\section{Obtaining thickness and refractive index from the reflection coefficient}\label{sec-dns}

To obtain the thickness of the sample, we rewrite equation~\ref{eq-s1} as
\eq{e^{2ikd\ns} & = & -\frac{\ns+\nw}{\ns-\nw}\,\frac{\ng-\ns-(\ng+\ns)s}{\ng+\ns-(\ng-\ns)s},\nonumber}
whereby
\eq{d & = & \frac{i}{2k\ns}\,\logg{}{-\frac{\ns-\nw}{\ns+\nw}\,\frac{\ng+\ns-(\ng-\ns)s}{\ng-\ns-(\ng+\ns)s}},\label{eq-d}}
where $\text{log}$ denotes the complex logarithm. It is now evident that this expression has an infinite number of values and that choosing one equates to choosing a logarithm branch. This is why the pattern seen on figure~\ref{fig-1-000} is periodic in $d$; its period is $2\pi/2k\ns$, as is evident from equation~\ref{eq-d}.

\begin{figure}[b!]
\begin{center}
\hrule\ \\\ \\
\begin{minipage}[m]{\figwidth}
\quad\includegraphics[width=\figwidth]{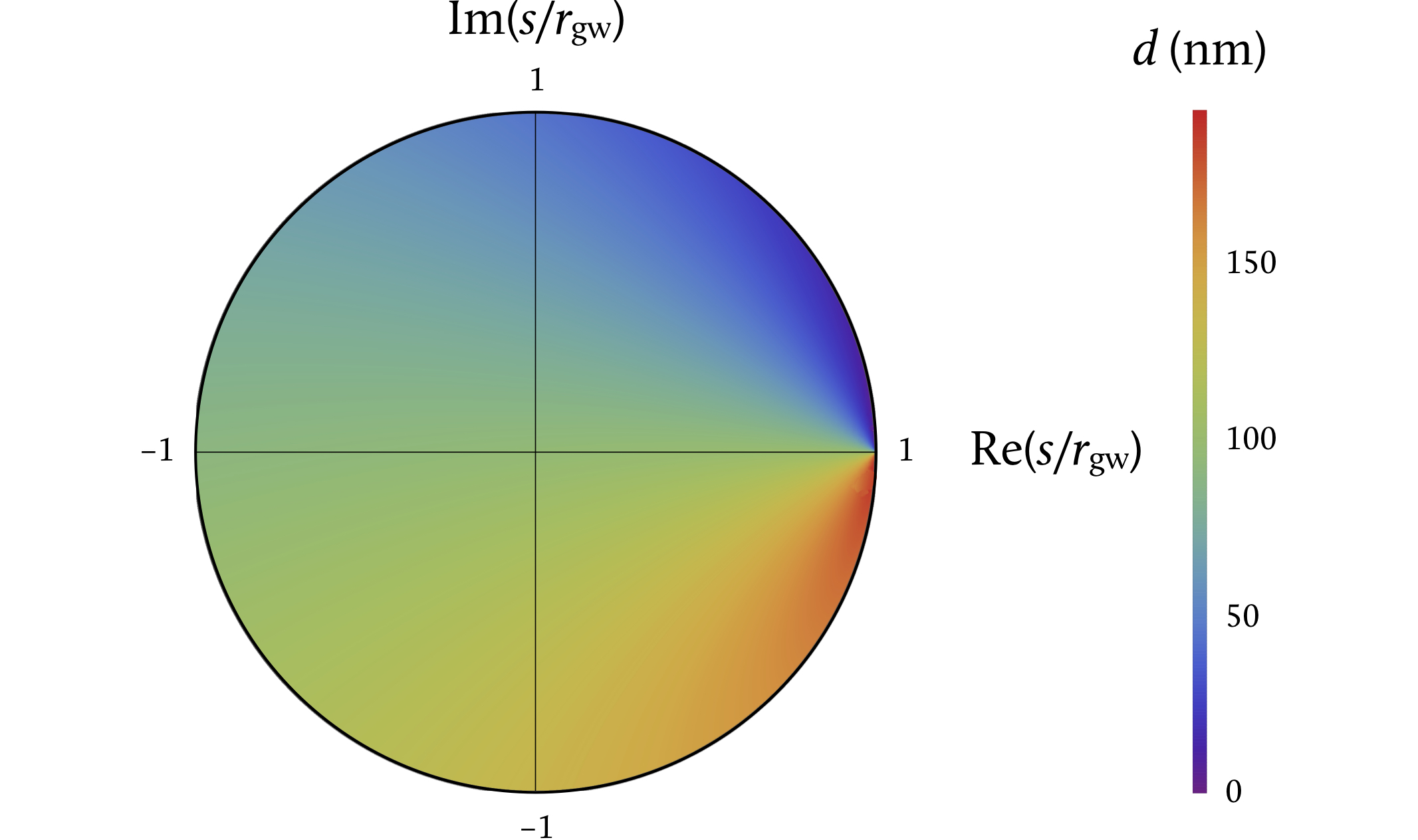}
\end{minipage}\\[2.5ex]
\begin{minipage}[m]{\figwidth}
\quad\includegraphics[width=\figwidth]{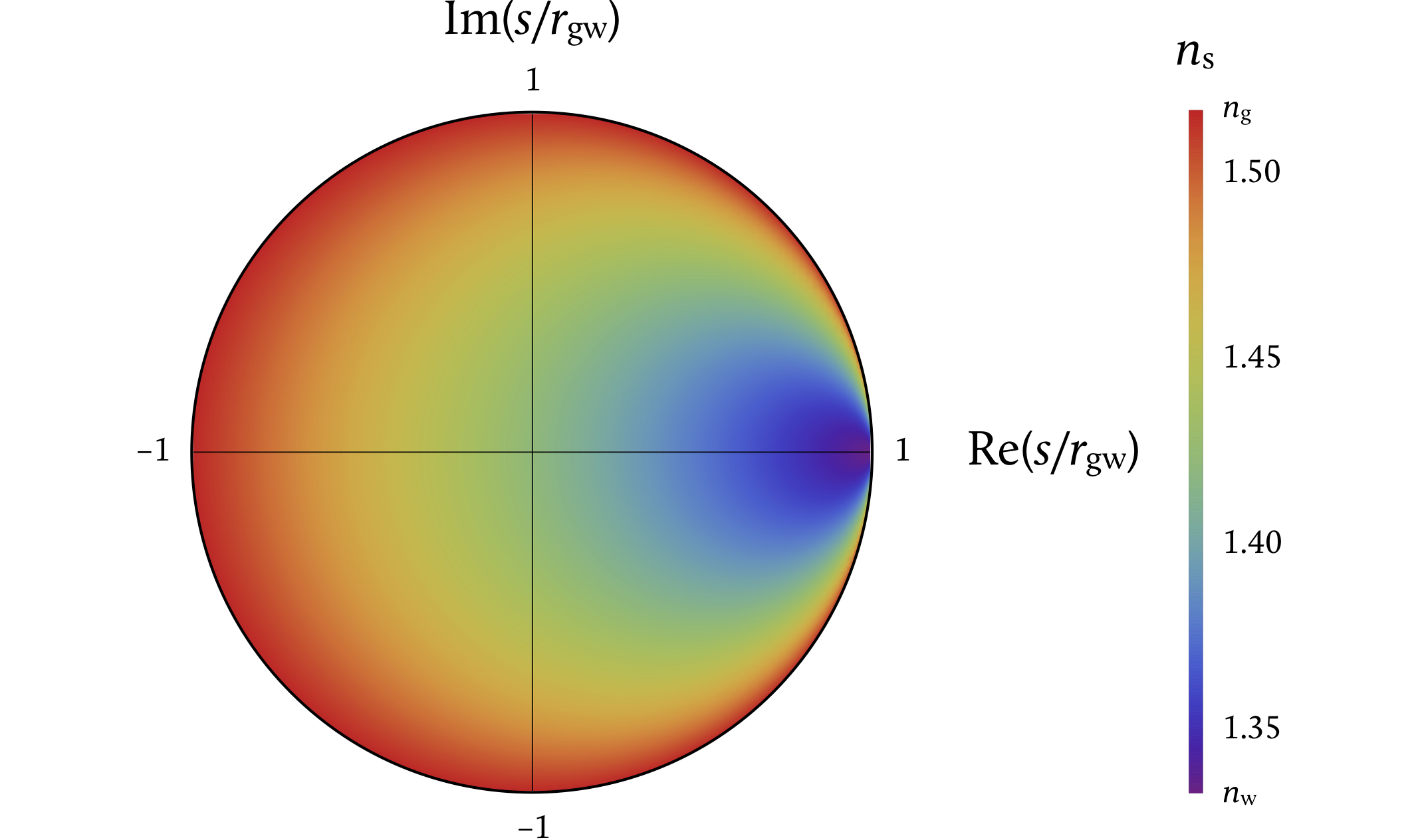}
\end{minipage}
\caption{Density graphs of the first period of $d$ (top) and $\ns$ (bottom) as functions of the real and imaginary parts of $s$ for normal incidence with $\nw=1.333$, $\ng=1.518$ and $\lambda=550$~nm.}\label{fig-1-dns-000}
\end{center}
\end{figure}

To obtain the refractive index of the sample, we note that the imaginary part of $d$ is
\eq{\Imag{d} & = & \frac{1}{2k\ns}\,\log{\left|\frac{\ns-\nw}{\ns+\nw}\,\frac{\ng+\ns-(\ng-\ns)s}{\ng-\ns-(\ng+\ns)s}\right|}.\nonumber}
But $d$ is the thickness of the sample and must thus be a real number. Therefore,
\eq{\left(\frac{\ns-\nw}{\ns+\nw}\right)^2\frac{\ng+\ns-(\ng-\ns)s}{\ng-\ns-(\ng+\ns)s}\,\frac{\ng+\ns-(\ng-\ns)s^\ast}{\ng-\ns-(\ng+\ns)s^\ast} & = & 1.\nonumber}
From this expression we finally obtain
\eq{\ns & = & \sqrt{\nw\ng\,\frac{\ng-\nw-2\ng\Real{s}+(\ng+\nw)|s|^2}{\ng-\nw-2\nw\Real{s}-(\ng+\nw)|s|^2}}.\label{eq-ns}}

As noted in section~\ref{sec-s}, $|s/\rgw|\leq1$ as long as $\nw\leq\ns\leq\ng$. This is expected because in this refractive index range the difference between $\ng$ and $\nw$ is greater than the difference between $\ng$ and $\ns$, which results in the glass-water interface being more reflective than the glass-material interface. Taking this into account, we may graph $d$ and $\ns$ in the unit circle of the $\Real{s/\rgw}\times\Imag{s/\rgw}$ plane (figure~\ref{fig-1-dns-000}).

\section{Remarks on some mathematical properties of the reflection coefficient}\label{sec-rphi}

Substituting $s=0$ in equation~\ref{eq-ns}, we see immediately that the zeros of $s$ occur at $\ns=\sqrt{\nw\ng}$.

Substituting $\ns=\sqrt{\nw\ng}$ and $s=0$ in equation~\ref{eq-s1}, rearranging terms and writing $e^{i\psi}$ as $\cos{\psi}+i\,\sin{\psi}$, we obtain
\eq{\cos{2kd\sqrt{\nw\ng}}+i\,\sin{2kd\sqrt{\nw\ng}} & = & -\frac{\sqrt{\nw\ng}-\nw}{\sqrt{\nw\ng}+\nw}\,\frac{\ng+\sqrt{\nw\ng}}{\ng-\sqrt{\nw\ng}} \tab = \tab -1,\nonumber}
whereby
\eq{d & = & \frac{(2N-1)\pi}{2k\sqrt{\nw\ng}}\nonumber}
with $N\in\mathbb{N}$. This means that the zeros occur at the value of $d$ which is exactly at the centre of each period of $s$.

Thus, $d=(2N-1)\lambda/4\sqrt{\nw\ng}$ and $\ns=\sqrt{\nw\ng}$ are the conditions the layer must have in order to constitute a perfect antireflective coating for light of wavelength $\lambda$. Conversely, having $d=N\lambda/2\ns$ for any value of $\ns$ is, in terms of reflectivity, equivalent to not having a layer at all.

\begin{figure}[b!]
\begin{center}
\hrule\ \\\ \\
\scalebox{1}{
\begin{tikzpicture}
\pgfplotsset{every axis legend/.append style={at={(1.5,1)},anchor=north east}}
\begin{axis}[width=\graphwidth,height=\graphheight,legend cell align=left,legend style={draw=none,row sep=10pt},xlabel=$d$ (nm),xmin=0,xmax=650,ylabel=$\arg{s}$,ymin=-3.5,ymax=3.5,scaled ticks=false,yticklabel style={/pgf/number format/.cd,fixed,fixed zerofill,precision=0},xtick={0,100,200,300,400,500,600},ytick={-3.142,-1.571,0.01,1.571,3.142},yticklabels={$-\pi$,$-\displaystyle\frac{\pi}{2}$,$0$,$\displaystyle\frac{\pi}{2}$,$\pi$},axis x line=middle,x label style={at={(current axis.right of origin)},anchor=west},axis y line=center,y label style={at={(current axis.above origin)},anchor=south}]
\addplot[smooth,thick,red] file[x index=0,y index=1] {dat-nsargs-1.dat};
\addplot[smooth,thick,orange] file[x index=0,y index=1] {dat-nsargs-2.dat};
\addplot[smooth,thick,yelloww] file[x index=0,y index=1] {dat-nsargs-3.dat};
\addplot[smooth,thick,orange] file[x index=0,y index=1] {dat-nsargs-4.dat};
\addplot[smooth,thick,orange] file[x index=0,y index=1] {dat-nsargs-5.dat};
\addplot[smooth,thick,orange] file[x index=0,y index=1] {dat-nsargs-6.dat};
\addplot[thick,yelloww] file[x index=0,y index=1] {dat-nsargs-7.dat};
\addplot[thick,yelloww] file[x index=0,y index=1] {dat-nsargs-8.dat};
\addplot[thick,yelloww] file[x index=0,y index=1] {dat-nsargs-9.dat};
\addplot[black,dashed] coordinates {
(0,-3.142)
(600,-3.142)};
\addplot[black,dashed] coordinates {
(0,-1.571)
(600,-1.571)};
\addplot[black,dashed] coordinates {
(0,1.571)
(600,1.571)};
\addplot[black,dashed] coordinates {
(0,3.142)
(600,3.142)};
\legend{$\ns=1.38$,$\ns=\sqrt{\nw\ng}$,$\ns=1.47$}
\end{axis}
\end{tikzpicture}}
\caption{The effect of $\ns$ on the range of $\arg{s}$. If $\ns<\sqrt{\nw\ng}$ (red curve), $\arg{s}$ can only take small values; if $\ns=\sqrt{\nw\ng}$ (orange curve), $\arg{s}$ can take all values between $-\pi/2$ and $\pi/2$; if $\ns>\sqrt{\nw\ng}$ (yellow curve), $\arg{s}$ can take any value. Here, $\nw$, $\ng$ anad $\lambda$ are as in previous figures.}
\label{fig-nsargs}
\end{center}
\end{figure}
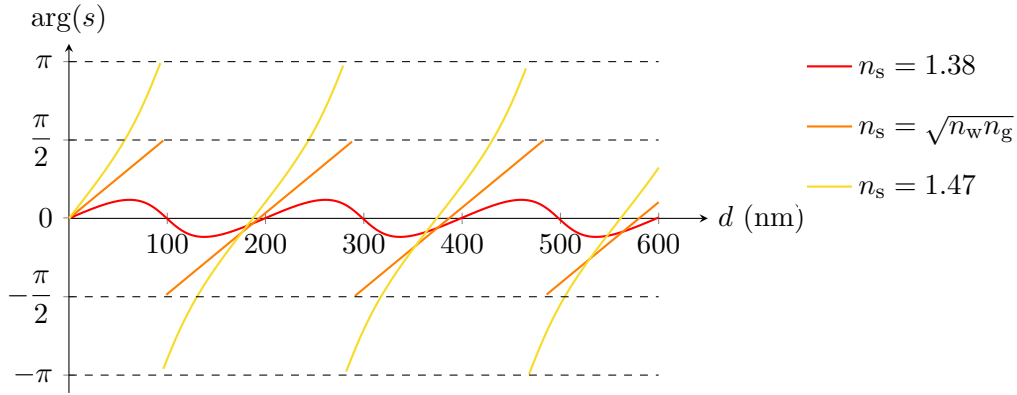

Let us now recall that there is a value of $\ns$ below which $\Real{s}>0$. To find this value, we first multiply and divide $s$ in equation~\ref{eq-s1} by the conjugate of the denominator of the right-hand-term to obtain
\eq{\frac{s}{2} & = & \frac{(\ng\.^2-\ns\.^2)(\ns\.^2+\nw\.^2)+(\ns\.^2-\nw\.^2)\!\left((\ng\.^2+\ns\.^2)\cos{\psi}+2i\ng\ns\sin{\psi}\right)}{(\ng-\ns)^2(\ns-\nw)^2+(\ng+\ns)^2(\ns+\nw)^2+2(\ng\.^2-\ns\.^2)(\ns\.^2-\nw\.^2)\cos{\psi}},\nonumber}
where $\psi=2kd\ns$. The denominator is always positive because it is the product of a complex number and its conjugate, so the values of $\ns$ for which the real part of the denominator is negative are exactly those values for which $\Real{s}<0$.

We thus have the condition
\eq{\cos{2kd\ns} & < & -\frac{(\ng\.^2-\ns\.^2)(\ns\.^2+\nw\.^2)}{(\ng\.^2+\ns\.^2)(\ns\.^2-\nw\.^2)},\nonumber}
which requires
\eq{\frac{(\ng\.^2-\ns\.^2)(\ns\.^2+\nw\.^2)}{(\ng\.^2+\ns\.^2)(\ns\.^2-\nw\.^2)} & < & 1.\nonumber}

Now, this is a monotonically decreasing function of $\ns$ in the range $\nw<\ns<\ng$, as shown by the fact that
\eq{\pder{}{\ns}\!\left(\frac{(\ng\.^2-\ns\.^2)(\ns\.^2+\nw\.^2)}{(\ng\.^2+\ns\.^2)(\ns\.^2-\nw\.^2)}\right) & = & -4\ns\,\frac{(\ng\.^2-\nw\.^2)(\nw\.^2\ng\.^2+\ns\.^4)}{(\ng\.^2+\ns\.^2)^2(\ns\.^2-\nw\.^2)^2} \tab < \tab 0.\nonumber}
Therefore, the value of $\ns$ for which it equals $1$ is the value of $\ns$ below which $\Real{s}$ is necessarily non-negative. If we write $\ns=\sqrt{\nw\ng}$, we obtain
\eq{\frac{(\ng\.^2-\ns\.^2)(\ns\.^2+\nw\.^2)}{(\ng\.^2+\ns\.^2)(\ns\.^2-\nw\.^2)} & = & \frac{\nw\ng(\ng\.^2-\nw\.^2)}{\nw\ng(\ng\.^2+\nw\.^2)} \tab = \tab 1,\nonumber}
so $\sqrt{\nw\ng}$ is the value we seek (figure~\ref{fig-nsargs}), as already suggested by figure~\ref{fig-1-000}.

\section{Non-normal incidence}\label{sec-NA}

For non-normal incidence, the Fresnel reflection and transmission coefficients are
\eq{r_{jk}^\parallel & = & \frac{n_j\cos{\theta_j}-n_k\cos{\theta_k}}{n_j\cos{\theta_j}+n_k\cos{\theta_k}},\nonumber\\
t_{jk}^\parallel & = & \frac{2n_j\cos{\theta_j}}{n_j\cos{\theta_j}+n_k\cos{\theta_k}}\nonumber}
for light polarised parallel to the plane of incidence and
\eq{r_{jk}^\perp & = & \frac{n_k\cos{\theta_j}-n_j\cos{\theta_k}}{n_k\cos{\theta_j}+n_j\cos{\theta_k}},\nonumber\\
t_{jk}^\perp & = & \frac{2n_j\cos{\theta_j}}{n_k\cos{\theta_j}+n_j\cos{\theta_k}}\nonumber}
for light polarised perpendicular to the plane of incidence, where $\theta_j$ is the angle of incidence and
\eq{\theta_k & = & \arcsin{\frac{n_k}{n_j}\,\sin{\theta_j}\!}\nonumber}
is the angle of transmission, given by Snell's law. Obviously, then, $s$ is a function of the angle of incidence as well as of $d$ and $\ns$. We will henceforth assume that $\theta_j$ is small enough and $n_k$ is close enough to $n_j$ to avoid total internal reflection when $n_j>n_k$.

If the incident light is circularly polarised, the parallel and perpendicular components have equal amplitude, so we have
\eq{s & = & \frac{s_\parallel+s_\perp}{2}.\nonumber}

\fig{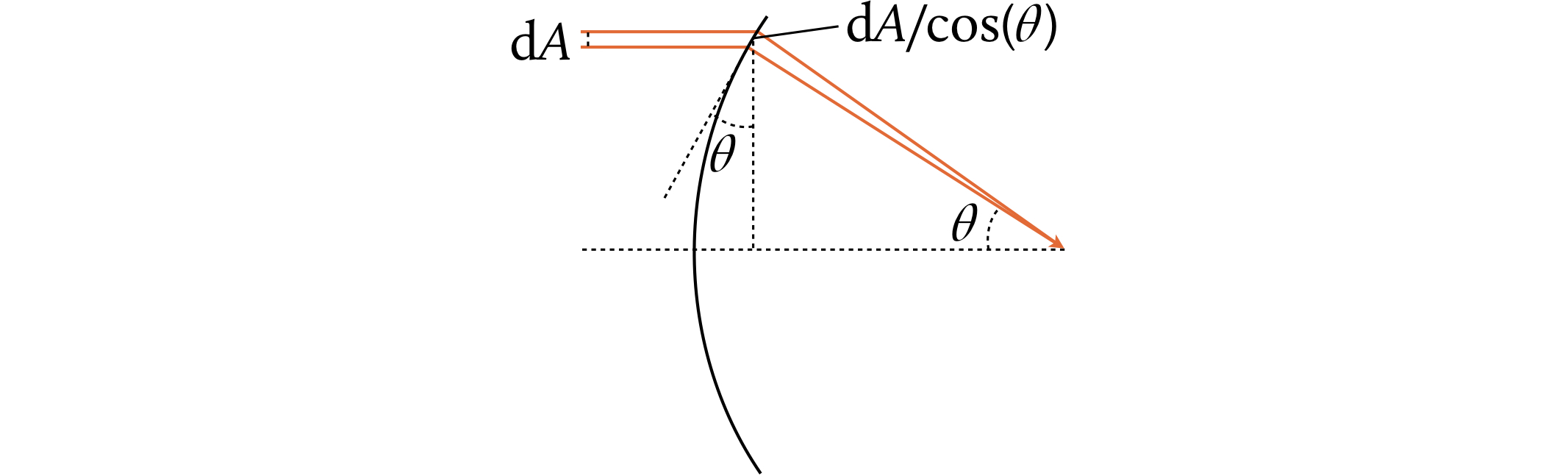}{\figwidth}{The projection of an area element $\text{d}A$ of the incident beam onto the reference sphere of an aplanatic objective is $\text{d}A/\cos{\theta}$.}{fig-aplanatic}

For a distribution $\mathcal{P}(\theta,\varphi)$ of angles of incidence, the reflectometry signal $S$, which is given by the interference between a beam that interacts with the sample and a beam that does not and thus contains the angular distribution of both beams, must be averaged over all possible angles. This average is given by
\eq{S & = & \frac{1}{2\pi}\,\integral{0}{2\pi}{\integral{0}{\thetamax}{\mathcal{P}(\theta,\varphi)\,s(d,\ns,\theta)\,\cos{\theta}\sin{\theta}}{\theta}}{\varphi},\nonumber}
where $\thetamax$ is the maximum incidence angle of the light incident on the sample; if $\thetamax$ is determined by the numerical aperture $\text{NA}$ of a microscope objective, for instance, then
\eq{\thetamax & = & \arcsin{\frac{\text{NA}}{\ng}}.\nonumber}
The cosine in the integral comes from assuming the objective in question is aplanatic; the projection of an area element $\text{d}A$ of the incident interfered beams onto the aplanatic lens reference sphere is $\text{d}A/\cos{\theta}$ (figure~\ref{fig-aplanatic}).\cite{ref-iRef-NovotnyPONO}

It should be noted that the critical angle, the angle at which total internal reflection occurs, is
\eq{\thetac & = & \arcsin{\frac{\nw}{\ng}} \tab = \tab 61.42\ ^\circ\nonumber}
for the glass-water interface and even higher for the glass-layer and layer-water interfaces if the layer has a refractive index between $\nw$ and $\ng$. For an objective with a numerical aperture of $1.27$, for example, $\thetamax=56.79$~$^\circ$, so total internal reflection is not a problem at the glass-layer and glass-water interfaces. Light travelling through the glass at an angle $\theta\leq\thetamax$ will be transmitted into the layer at an angle
\eq{\thetas & = & \arcsin{\frac{\ng}{\ns}\,\sin{\theta}},\nonumber}
so at the layer-interface we have
\eq{\thetas-\thetac & = & \arcsin{\frac{\ng}{\ns}\,\sin{\theta}}-\arcsin{\frac{\nw}{\ns}},\nonumber}
which is a monotonically increasing function of both $\ns$ and $\theta$ but is negative even for the highest value of $\ns$ we are considering, $\ng$; therefore, there will be no total internal reflection at the layer-water interface either.

\begin{figure}[b!]
\begin{center}
\hrule\ \\\ \\
\begin{minipage}[m]{\figwidth}
\qquad\includegraphics[width=\figwidth]{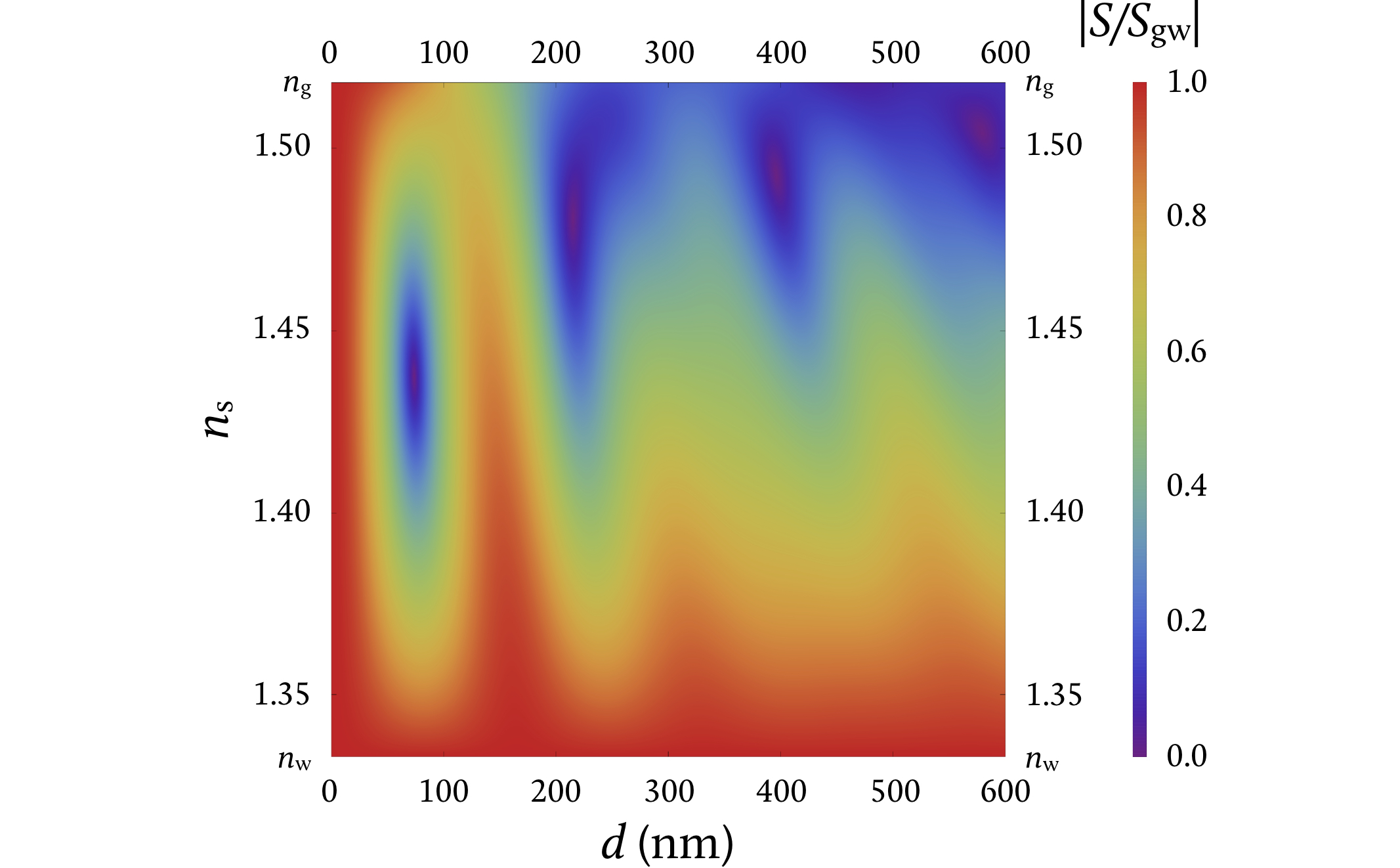}
\end{minipage}\\[2.5ex]
\begin{minipage}[m]{\figwidth}
\qquad\includegraphics[width=\figwidth]{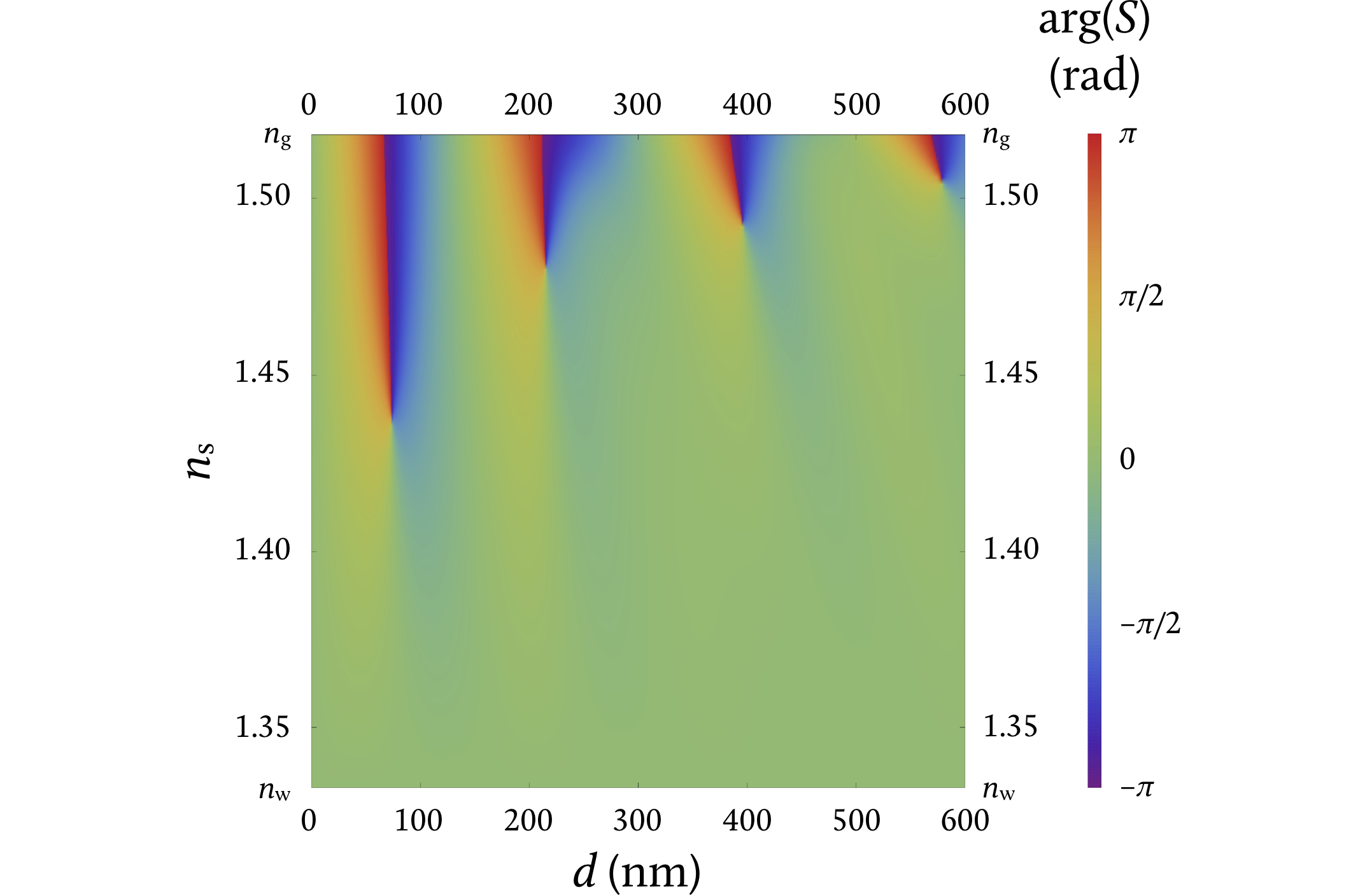}
\end{minipage}
\caption{Density graphs of $|S/S_\text{gw}|$ (top) and $\arg{S}$ (bottom) as functions of $d$ and $\ns$ for the case in which both beams are gaussian; the beam that interacts with the sample emerges from (and is then reflected back through) an aplanatic objective with numerical aperture $1.27$ and fill factor $1$; and $\nw=1.333$, $\ng=1.518$ and $\lambda=550$~nm.}\label{fig-1-127}
\end{center}
\end{figure}

In the case in which the both beams are gaussian and have the same angular distribution, the angular dependence of the detected signal is a function of only $\theta$ and is given by\cite{ref-iRef-NovotnyPONO}
\eq{\mathcal{P}(\theta) & = & e^{-\zeta^2\,\frac{\sinn{2}{\theta}}{\sinn{2}{\thetamax}}},\nonumber}
where $\zeta$ is the objective fill factor. This turns the signal into
\eq{S & = & \integral{0}{\thetamax}{e^{-\zeta^2\,\frac{\sinn{2}{\theta}}{\sinn{2}{\thetamax}}}\,\frac{s_\parallel+s_\perp}{2}\,\cos{\theta}\sin{\theta}}{\theta}.\label{eq-s1avg}}
In this case, the expression for $S(d,\ns)$ can no longer be solved analytically.

\begin{figure}[b!]
\begin{center}
\hrule\ \\\ \\
\begin{minipage}[m]{\figwidth}
\quad\includegraphics[width=\figwidth]{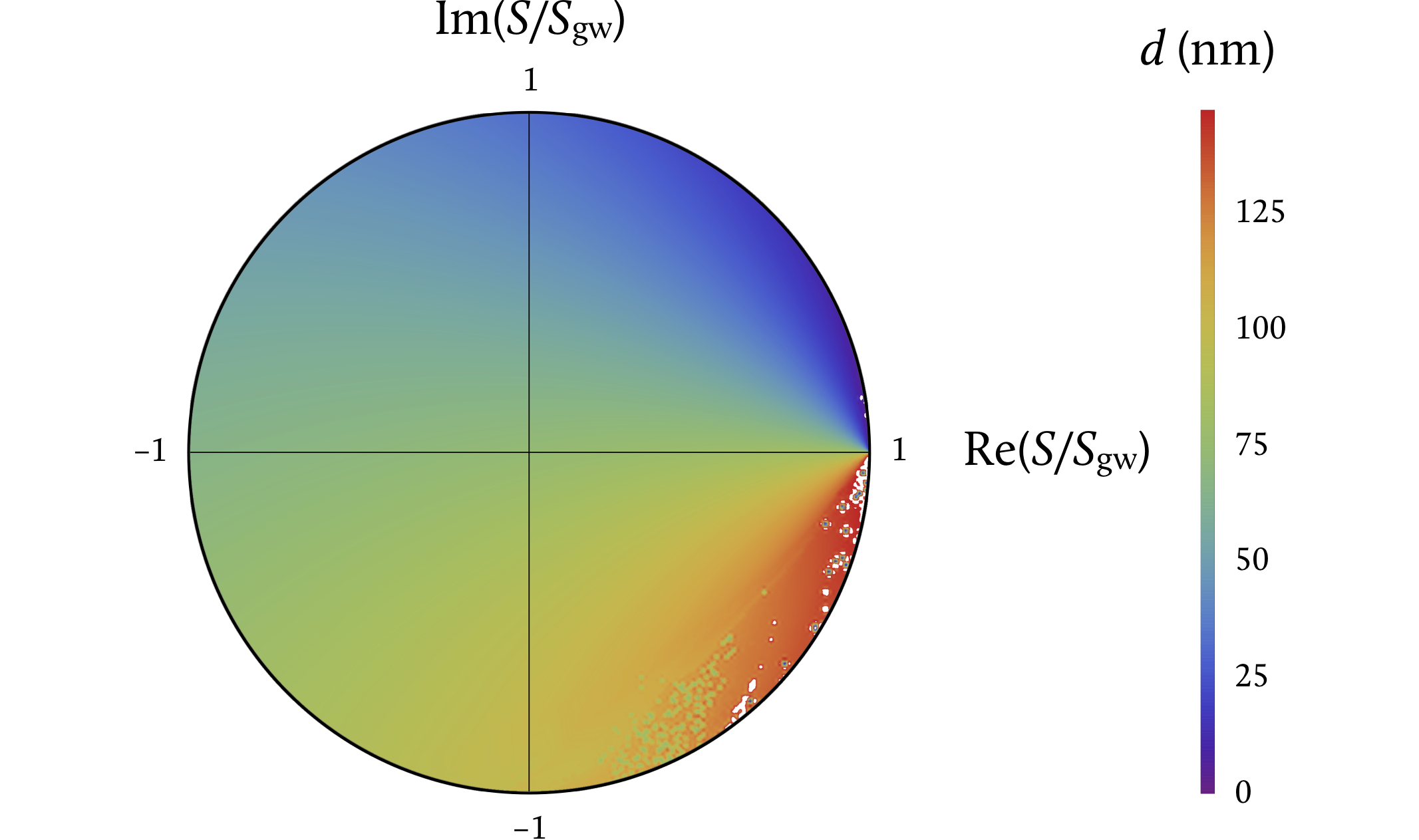}
\end{minipage}\\[2.5ex]
\begin{minipage}[m]{\figwidth}
\quad\includegraphics[width=\figwidth]{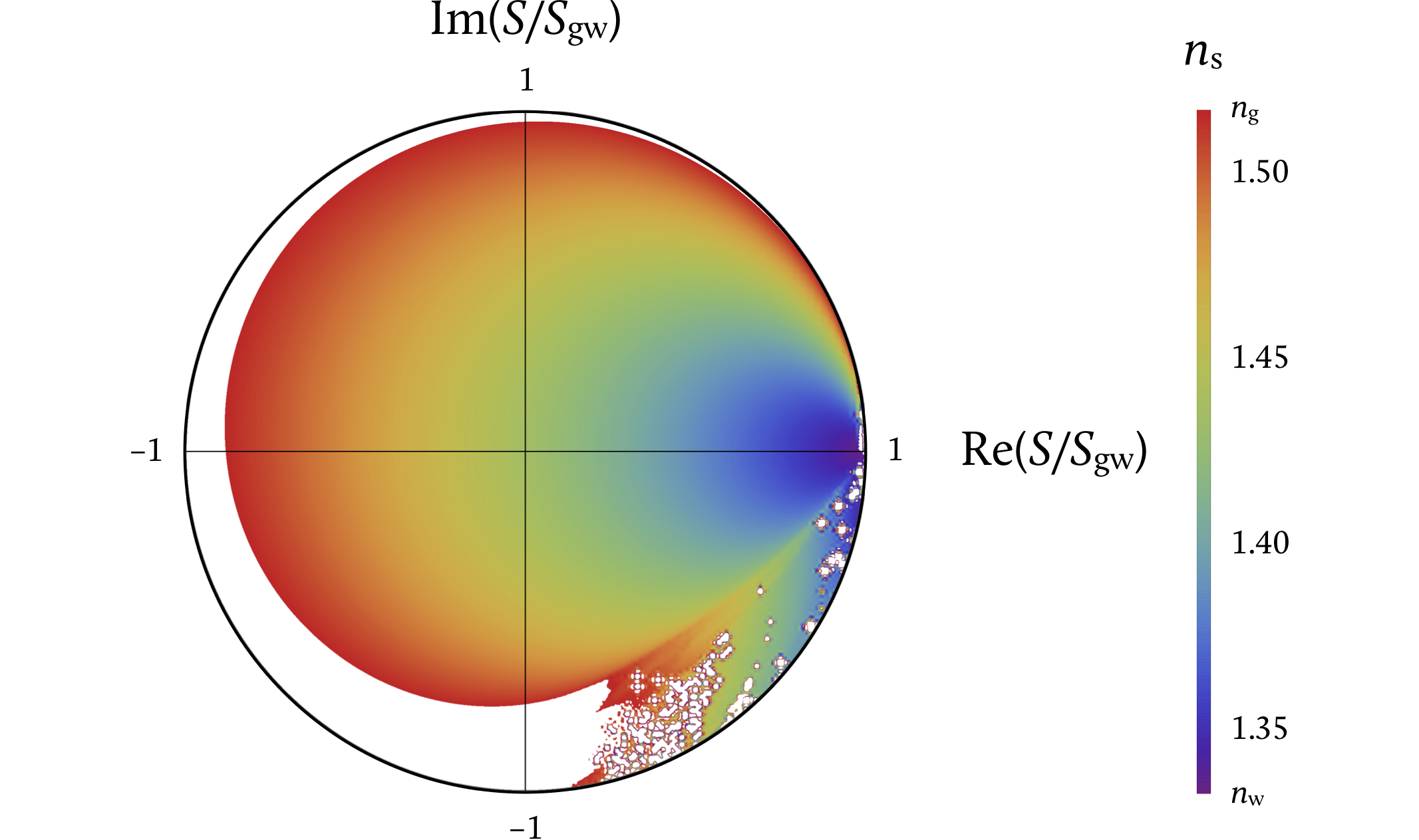}
\end{minipage}
\caption{Density graphs of the first repetition of $d$ (top) and $\ns$ (bottom) as functions of the real and imaginary parts of $S/S_\text{gw}$ for the case in which both beams are gaussian; the beam that interacts with the sample emerges from (and is then reflected back through) an objective with numerical aperture $1.27$ and fill factor $1$; and $\nw=1.333$, $\ng=1.518$ and $\lambda=550$~nm. The irregularly coloured section in the fourth quadrant of the $\ns$ graph is an artefact of part of the second repetition having been included in the calculation of $\ns$, which occurred because a rectangular section of $(d,\ns)$ space was taken and the repetitions are not rectangular.}\label{fig-1-dns-127}
\end{center}
\end{figure}

For a small value of $\thetamax$ (i.e.~for a small NA), the relative reflection coefficient is only slightly deformed with respect to that observed for normal incidence; for larger values of $\thetamax$, the pattern is no longer periodic (although it retains a partially repetitive behaviour) and its deformation becomes more evident (compare figures~\ref{fig-1-000} and~\ref{fig-1-127}).

Figure~\ref{fig-1-127} shows the amplitude and phase of the reflectometry signal, $S$, normalised with respect to the no-layer signal, $S_\text{gw}$ (given by replacing $s$ with $\rgw$ in equation~\ref{eq-s1avg}), for gaussian beams, an objective with a numerical aperture of $1.27$ and a fill factor of $1$, and all other parameters as before. With these parameters, $S_\text{gw}\approx-0.0063$.

It is immediately evident that $|S/S_\text{gw}|\leq1$ as long as $\nw\leq\ns\leq\ng$. We may thus graph $d$ and $\ns$ in the unit circle of the $\Real{S/S_\text{gw}}\times\Imag{S/S_\text{gw}}$ plane. To do so, we must take into account only one repetition (as mentioned earlier, the behaviour of $S$ for non-normal incidence is still partially repetitive, although the values of $\ns$ for which $S$ reaches the same value in different repetitions are different from each other).

Figure~\ref{fig-1-dns-127} shows $d$ and $\ns$ as functions of $\Real{S/S_\text{gw}}$ and $\Imag{S/S_\text{gw}}$ for a gaussian beam emerging from an objective with a numerical aperture of $1.27$ and a fill factor of $1$. Comparing this to figure~\ref{fig-1-dns-000}, it is immediately evident that, while the entire $\left(|S/S_\text{gw}|,\arg{S}\right)\in[0,1]\times[-\pi,\pi]$ space yields values of $\ns$ between $\nw$ and $\ng$ in the case of normal incidence, this is not so in the case of large numerical aperture, as expected from careful examination of the bottom, left and top borders of figures~\ref{fig-1-000} and~\ref{fig-1-127}: whereas in the case of normal incidence $\arg{S}$ takes all possible values between $-\pi$ and $\pi$ along the top border (where $\ns=\ng$ and $|S/S_\text{gw}|=1$), a nonzero numerical aperture causes $|S/S_\text{gw}|$ to no longer equal $1$ along the top border, which is the only one of these borders along which $\arg{S}$ takes nonzero values; for a nonzero numerical aperture, $\arg{S}=0$ all along the left ($d=0$) and bottom ($\ns=\nw$) borders, which are the only places where $|S/S_\text{gw}|=1$; this translates into figure~\ref{fig-1-dns-127} as the white region where $|S/S_\text{gw}|\lesssim1$ and $\arg{S}\neq0$ simultaneously.

\section{Reflection by multiple layers}\label{sec-multiplelayers}

\subsection{Two layers}

\fig{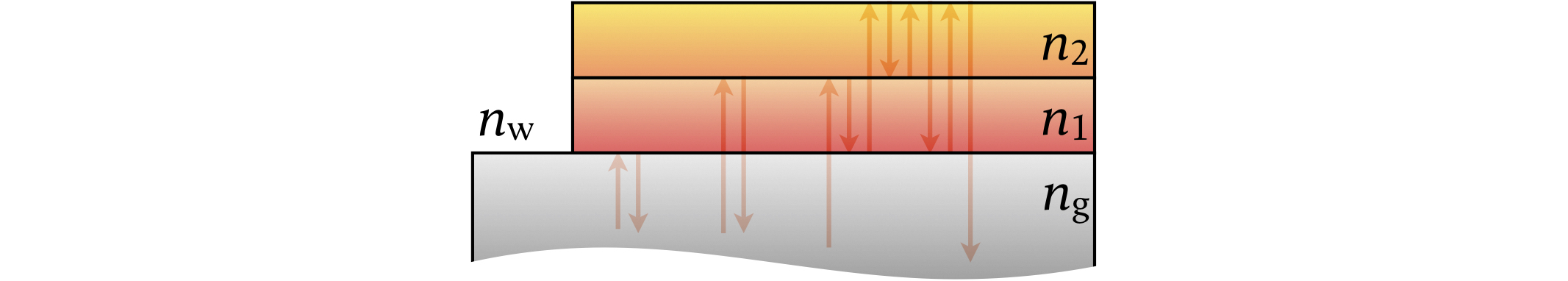}{\figwidth}{Reflection from a thin sample consisting of two layers of different materials.}{fig-thinsample2}

If the sample consists of two layers, each with its own thickness and refractive index (figure~\ref{fig-thinsample2}), the situation becomes more complicated. Let the angle of incidence be denoted by $\theta$, as before, and let $\theta_1$ and $\theta_2$ be the transmission angles in the first and second layers, respectively, given by Snell's law. Let us further denote the properties of the materials by $d_1$, $n_1$, $d_2$ and $n_2$. We now have the following possibilities:

The light may be reflected at the interface between the glass and the first material and never enter the sample, as before. This contributes a term $s_0=r_{\text{g}1}$.

The transmitted light may enter the first material and then be reflected any number $\ell+1$ of times at the interface between the two materials and $\ell$ times at the interface between the first material and the glass before finally being transmitted back through the glass. This contributes a term similar to the one in the 1-layer case:
\eq{s_1 & = & t_{\text{g}1}r_{12}t_{1\text{g}}e^{2ik\,\frac{d_1 n_1}{\cos{\theta_1}}}\sum_{\ell=0}^\infty r_{1\text{g}}\.^\ell r_{12}\.^\ell e^{2\ell ik\,\frac{d_1 n_1}{\cos{\theta_1}}}.\nonumber}

Finally, the light may be transmitted through the interface between the two materials and enter the second layer (after any number of reflections within the first layer). This light can be either transmitted into the water, in which case it does not contribute to the reflection coefficient, or reflected back into the second layer and then transmitted back through the first layer and into the glass (after any number of reflections within either material or within both materials). Since we have different kinds of reflections, there are multiple ways of ordering them, so a multiplicity factor must be included in the two-layer term. If we have $j$ reflections within the first bilayer, $\ell$ reflections within both bilayers and $m$ reflections within the second bilayer, there are $(j+\ell+m)!/j!\ell!m!$ different ways of ordering them. The term contributed by this case is then
\eq{s_2 & = & t_{\text{g}1}t_{12}r_{2\text{w}}t_{21}t_{1\text{g}}e^{2ik\left(\frac{d_1 n_1}{\cos{\theta_1}}+\frac{d_2 n_2}{\cos{\theta_2}}\right)}\times\vphantom{\sum_{j=0}^\infty\frac{(j+\ell+m)!}{j!\ell!m!}}\nonumber\\
& & \times\sum_{j=0}^\infty\sum_{\ell=0}^\infty\sum_{m=0}^\infty\frac{(j+\ell+m)!}{j!\ell!m!}\,r_{1\text{g}}\.^j r_{12}\.^j e^{2jik\,\frac{d_1 n_1}{\cos{\theta_1}}}\times\vphantom{e^{2ik\left(\frac{d_1 n_1}{\cos{\theta_1}}+\frac{d_2 n_2}{\cos{\theta_2}}\right)}}\nonumber\\
& & \phantom{\times\sum_{j=0}^\infty\sum_{\ell=0}^\infty\sum_{m=0}^\infty}\times r_{1\text{g}}\.^\ell t_{12}\.^\ell r_{2\text{w}}\.^\ell t_{21}\.^\ell e^{2\ell ik\left(\frac{d_1 n_1}{\cos{\theta_1}}+\frac{d_2 n_2}{\cos{\theta_2}}\right)}r_{21}\.^m r_{2\text{w}}\.^m e^{2mik\,\frac{d_2 n_2}{\cos{\theta_2}}}.\vphantom{e^{2ik\left(\frac{d_1 n_1}{\cos{\theta_1}}+\frac{d_2 n_2}{\cos{\theta_2}}\right)}\sum_{j=0}^\infty\frac{(j+\ell+m)!}{j!\ell!m!}}\nonumber}

The total reflection coefficient is, of course, $s=s_0+s_1+s_2$, which simplifies to
\eq{s & = & r_{\text{g}1}+\frac{t_{\text{g}1}r_{12}t_{1\text{g}}e^{2ik\,\frac{d_1 n_1}{\cos{\theta_1}}}}{1-r_{1\text{g}}r_{12}e^{2ik\,\frac{d_1 n_1}{\cos{\theta_1}}}}\vphantom{\frac{t_{\text{g}1}t_{12}r_{2\text{w}}t_{21}t_{1\text{g}}e^{2ik\left(\frac{d_1 n_1}{\cos{\theta_1}}+\frac{d_2 n_2}{\cos{\theta_2}}\right)}}{1-r_{1\text{g}}r_{12}e^{2ik\,\frac{d_1 n_1}{\cos{\theta_1}}}-r_{1\text{g}}t_{12}r_{2\text{w}}t_{21}e^{2ik\left(\frac{d_1 n_1}{\cos{\theta_1}}+\frac{d_2 n_2}{\cos{theta_2}}\right)}-r_{21}r_{2\text{w}}e^{2ik\,\frac{d_2 n_2}{\cos{\theta_2}}}}}\nonumber\\
& & +\ \frac{t_{\text{g}1}t_{12}r_{2\text{w}}t_{21}t_{1\text{g}}e^{2ik\left(\frac{d_1 n_1}{\cos{\theta_1}}+\frac{d_2 n_2}{\cos{\theta_2}}\right)}}{1-r_{1\text{g}}r_{12}e^{2ik\,\frac{d_1 n_1}{\cos{\theta_1}}}-r_{1\text{g}}t_{12}r_{2\text{w}}t_{21}e^{2ik\left(\frac{d_1 n_1}{\cos{\theta_1}}+\frac{d_2 n_2}{\cos{\theta_2}}\right)}-r_{21}r_{2\text{w}}e^{2ik\,\frac{d_2 n_2}{\cos{\theta_2}}}}.\qquad\label{eq-s2}}

As a sanity check, we may set $n_2=n_1$ (which will turn the Fresnel coefficients for the interface between the two layers into $r_{12}=r_{21}=0$ and $t_{12}=t_{21}=1$). This yields
\eq{s & = & r_{\text{g}1}+\frac{t_{\text{g}1}r_{1\text{w}}t_{1\text{g}}e^{2ik\,\frac{(d_1+d_2)n_1}{\cos{\theta_1}}}}{1-r_{1\text{g}}r_{1\text{w}}e^{2ik\,\frac{(d_1+d_2)n_1}{\cos{\theta_1}}}},\nonumber}
which (except for the cosines, which arise from non-normal incidence) is identical to equation~\ref{eq-s1} with $d=d_1+d_2$ and $\ns=n_1$, as is expected. Similarly, substituting $n_2=\nw$ results in $r_{2\text{w}}=0$ and thus
\eq{s & = & r_{\text{g}1}+\frac{t_{\text{g}1}r_{1\text{w}}t_{1\text{g}}e^{2ik\,\frac{d_1 n_1}{\cos{\theta_1}}}}{1-r_{1\text{g}}r_{1\text{w}}e^{2ik\,\frac{d_1 n_1}{\cos{\theta_1}}}},\nonumber}
which is again equation~\ref{eq-s1} with $d=d_1$ and $\ns=n_1$. Notably, substituting $d_2=0$ does not reduce equation~\ref{eq-s2} to the 1-layer case; this is because $r_{1\text{w}}$ (which would be the new reflection coefficient at the top of the first layer) is not equal to $t_{12}r_{2\text{w}}t_{21}$.

In the single-reflection approximation, equation~\ref{eq-s2} becomes
\eq{s & \approx & r_{\text{g}1}+t_{\text{g}1}r_{12}t_{1\text{g}}e^{2ik\,\frac{d_1 n_1}{\cos{\theta_1}}}+t_{\text{g}1}t_{12}r_{2\text{w}}t_{21}t_{1\text{g}}e^{2ik\left(\frac{d_1 n_1}{\cos{\theta_1}}+\frac{d_2 n_2}{\cos{\theta_2}}\right)}.\nonumber}

\subsection{An arbitrary number of layers}

For $N$ layers with thicknesses $d_1,\ldots,d_N$ and refractive indices $n_1,\ldots,n_N$, we may separate the problem into $N+1$ partial reflection coefficients $s_0,\ldots,s_N$, each taking one more layer than the previous one, as we did for two layers. We may obtain a generalisable expression if we assign the index $0$ to the glass slide and and the index $N+1$ to the water.

We first define the symbol $\Xi$, which we will use to denote nested sums, as follows:
\eq{\stksum_{\ell,m,M}^{q,Q} & \equiv & \sum_{\ell_{q,q}=m}^M\sum_{\ell_{q,q+1}=m}^M\cdots\sum_{\ell_{q,Q}=m}^M\sum_{\ell_{q+1,q+1}=m}^M\cdots\sum_{\ell_{q+1,Q}=m}^M\cdots\sum_{\ell_{Q,Q}=m}^M.\nonumber}
Here, the first index of $\ell$ runs from $q$ to $Q$ and the second one runs from the first one's value to $Q$, so there are $(Q-q+1)(Q-q+2)/2$ sums. For example, for any function $f$,
\eq{\stksum_{\ell,0,\infty}^{1,3}f(\{\ell\}) & = & \sum_{\ell_{1,1}=0}^\infty\sum_{\ell_{1,2}=0}^\infty\sum_{\ell_{1,3}=0}^\infty\sum_{\ell_{2,2}=0}^\infty\sum_{\ell_{2,3}=0}^\infty\sum_{\ell_{3,3}=0}^\infty f(\ell_{1,1},\ell_{1,2},\ell_{1,3},\ell_{2,2},\ell_{2,3},\ell_{3,3}).\nonumber}

The $j$-th partial reflection coefficient is the combination of all possible reflections within each of the first $j$ layers, each pair of adjacent layers (with the corresponding transmission coefficients) within the first $j$ layers, and so on, taking all sets of $\ell$ adjacent layers with $1\leq\ell\leq j$ and remembering the multiplicity of each combination of reflections:
\eq{s_j & = & r_{j,j+1}\left(\prod_{\ell=1}^j t_{\ell-1,\ell}t_{\ell,\ell-1}\!\right)\!e^{2ik\sum\limits_{\ell=1}^j\frac{d_\ell n_\ell}{\cos{\theta_\ell}}}\times\vphantom{\frac{\left(\sum\limits_{\ell=1}^j\sum\limits_{q=\ell}^j m_{\ell,q}\!\right)!}{\prod\limits_{\ell=1}^j\prod\limits_{q=\ell}^j m_{\ell,q}!}}\nonumber\\
& & \times\stksum_{m,0,\infty}^{1,j}\left(\frac{\left(\sum\limits_{p=1}^j\sum\limits_{q=p}^j m_{p,q}\!\right)!}{\prod\limits_{p=1}^j\prod\limits_{q=p}^j m_{p,q}!}\,\prod_{p=1}^j\prod_{q=p}^j r_{p,p-1}\.^{m_{p,q}}r_{q,q+1}\.^{m_{p,q}}e^{2m_{p,q}ik\sum\limits_{\ell=p}^{q}\frac{d_\ell n_\ell}{\cos{\theta_\ell}}}\times\right.\nonumber\\
& & \phantom{\times\stksum_{m,0,\infty}^{1,j}\left(\frac{\left(\sum\limits_{p=1}^j\sum\limits_{q=p}^j m_{p,q}\!\right)!}{\prod\limits_{p=1}^j\prod\limits_{q=p}^j m_{p,q}!}\,\prod_{p=1}^j\prod_{q=p}^j\right.}\left.\times\prod_{\ell=p+1}^{q}t_{\ell-1,\ell}\.^{m_{p,q}}t_{\ell,\ell-1}\.^{m_{p,q}}\vphantom{\frac{\left(\sum\limits_{p=1}^j\sum\limits_{q=p}^j m_{p,q}\!\right)!}{\prod\limits_{p=1}^j\prod\limits_{q=p}^j m_{p,q}!}}\right).\nonumber}

The total reflection coefficient is thus
\eq{s & = & \sum_{j=0}^N r_{j,j+1}\left(\prod_{\ell=1}^j t_{\ell-1,\ell}t_{\ell,\ell-1}\!\right)\!e^{2ik\sum\limits_{\ell=1}^j\frac{d_\ell n_\ell}{\cos{\theta_\ell}}}\times\vphantom{\frac{\left(\sum\limits_{\ell=1}^j\sum\limits_{q=\ell}^j m_{\ell,q}\!\right)!}{\prod\limits_{\ell=1}^j\prod\limits_{q=\ell}^j m_{\ell,q}!}}\nonumber\\
& & \phantom{\sum_{j=0}^N}\times\stksum_{m,0,\infty}^{1,j}\left(\frac{\left(\sum\limits_{p=1}^j\sum\limits_{q=p}^j m_{p,q}\!\right)!}{\prod\limits_{p=1}^j\prod\limits_{q=p}^j m_{p,q}!}\,\prod_{p=1}^j\prod_{q=p}^j r_{p,p-1}\.^{m_{p,q}}r_{q,q+1}\.^{m_{p,q}}e^{2m_{p,q}ik\sum\limits_{\ell=p}^{q}\frac{d_\ell n_\ell}{\cos{\theta_\ell}}}\times\right.\qquad\nonumber\\
& & \phantom{\sum_{j=0}^N\times\stksum_{m,0,\infty}^{1,j}\left(\frac{\left(\sum\limits_{p=1}^j\sum\limits_{q=p}^j m_{p,q}\!\right)!}{\prod\limits_{p=1}^j\prod\limits_{q=p}^j m_{p,q}!}\,\prod_{p=1}^j\prod_{q=p}^j\right.}\left.\times\prod_{\ell=p+1}^{q}t_{\ell-1,\ell}\.^{m_{p,q}}t_{\ell,\ell-1}\.^{m_{p,q}}\vphantom{\frac{\left(\sum\limits_{p=1}^j\sum\limits_{q=p}^j m_{p,q}\!\right)!}{\prod\limits_{p=1}^j\prod\limits_{q=p}^j m_{p,q}!}}\right)\label{eq-s}}
with the understanding that, if $q>Q$, then
\eq{\sum_{\ell=q}^Q f & = & 0,\nonumber\\
\prod_{\ell=q}^Q f \tab = \tab \stksum_{\ell,m,M}^{q,Q} f & = & 1\nonumber}
for any function $f$.

Substitution of $N=0$, $N=1$ or $N=2$ into equation~\ref{eq-s} turns $s$ into the glass-water reflection coefficient ($\rgw$), the 1-layer reflection coefficient (equation~\ref{eq-s1}) or the 2-layer reflection coefficient (equation~\ref{eq-s2}), respectively.

In the single-reflection approximation, equation~\ref{eq-s} becomes
\eq{s & \approx & \sum_{j=0}^N r_{j,j+1}\left(\prod_{\ell=1}^j t_{\ell-1,\ell}t_{\ell,\ell-1}\right)e^{2ik\sum\limits_{\ell=1}^j\frac{d_\ell n_\ell}{\cos{\theta_\ell}}}.\nonumber}
Again, setting $N=1$, $N=1$ or $N=2$ turns this into $\rgw$ or the $1$- or $2$-layer single-reflection approximations, respectively.

\section{References}

\begingroup
\renewcommand{\section}[2]{}

\endgroup

\vfill

\end{document}